%% file: main.tex
\documentclass[final,3p,times,twocolumn]{elsarticle}

\usepackage{amssymb}
\usepackage{amsmath}
\usepackage{graphicx}
\usepackage{booktabs}
\usepackage{algorithm}
\usepackage{algorithmic}
\usepackage{subcaption}
\usepackage{xcolor}
\usepackage{tikz}
\usetikzlibrary{arrows.meta,positioning,patterns,fit,backgrounds,calc}
\usepackage{textcomp}
\usepackage[hyphens]{url}
\usepackage{hyperref}
\hypersetup{breaklinks=true}

\newcommand{\distest}{\textit{DistributedEstimator}}

\newcommand{\revised}[1]{#1}

\journal{Future Generation Computer Systems}

\begin{document}

\begin{frontmatter}

\title{DistributedEstimator: Distributed Training of Quantum Neural Networks via Circuit Cutting}

\input{author-desc}

\begin{abstract}
\revised{Circuit cutting decomposes a large quantum circuit into smaller subcircuits that are executed independently; the original circuit's expectation values are then recovered by classically combining the measured subcircuit outcomes.} While prior work characterises cutting overhead in terms of subcircuit counts and sampling complexity, its end-to-end impact on iterative, estimator-driven training pipelines remains insufficiently measured from a systems perspective. We propose \distest{}, a cut-aware estimator execution pipeline that treats circuit cutting as a staged distributed workload. Each estimator query is instrumented across four phases: partitioning, subexperiment generation, parallel execution, and classical reconstruction. Using logged runtime traces and learning outcomes on two binary classification workloads (Iris and MNIST), we quantify cutting overheads, scaling limits, and sensitivity to injected stragglers, and evaluate whether accuracy and robustness are preserved under matched training budgets. Our measurements reveal that reconstruction constitutes a dominant fraction of per-query time---reaching a median of 53\% and a 95th percentile of 58\% at three cuts---thereby bounding achievable speed-up under increased parallelism. Despite these overheads, test accuracy is fully preserved on Iris and maintained without systematic degradation on MNIST across all evaluated cut configurations. Robustness under Gaussian noise and FGSM perturbations is similarly preserved, with several cut configurations exhibiting comparable or improved robustness relative to the uncut baseline. The exponential growth of subexperiment counts with each additional cut ($\mathcal{O}(9^c)$ for CNOT-based decomposition) represents a fundamental computational barrier that limits practical experimentation to small qubit counts with current methods. These results establish that practical scaling of circuit cutting for learning workloads requires reducing and overlapping reconstruction, designing scheduling policies for barrier-dominated critical paths, and developing computationally efficient reconstruction strategies for larger qubit counts.
\end{abstract}



\end{frontmatter}

\input{section-introduction}
\input{section-background}
\input{section-related-work}
\input{section-system-model}
\input{section-methodology}
\input{section-evaluation}
\input{section-discussion}
\input{section-conclusion}
\input{section-credit}



\bibliographystyle{elsarticle-num}
\bibliography{references}

\end{document}

%% file: author-desc.tex
\author[melb]{Prabhjot Singh\corref{cor1}}
\ead{prabhjot.singh.1@student.unimelb.edu.au}
\author[melb]{Adel N. Toosi}
\author[melb]{Rajkumar Buyya}

\cortext[cor1]{Corresponding author.}

\affiliation[melb]{organization={Quantum Cloud Computing and Distributed Systems (qCLOUDS) Lab, \\
School of Computing and Information Systems, \\
The University of Melbourne},
            city={Melbourne},
            state={VIC},
            postcode={3010},
            country={Australia}}

%% file: section-introduction.tex

\section{Introduction}
\label{sec:introduction}

Hybrid quantum--classical workloads are increasingly shaped by systems constraints rather than purely algorithmic ones: the circuits and training loops of practical interest frequently exceed the width, depth, or reliability envelope of near-term devices, while still requiring repeated expectation estimation in an interactive optimisation loop~\cite{Preskill2018NISQ,Cerezo2021VQA,Endo2021QEM}. In this regime, execution is dominated by \emph{estimator-heavy} pipelines that combine many short quantum jobs with non-trivial classical orchestration, aggregation, and optimiser feedback~\cite{Schuld2019Gradients,Temme2017ErrorMitigation}. As a result, the end-to-end performance of quantum machine learning (QML) and variational training is governed not only by circuit latency, but also by how effectively we schedule, parallelise, and reconstruct large collections of sub-tasks under heterogeneous runtimes~\cite{DeanGhemawat2004MapReduce,Zaharia2008LATE,Sergeev2018Horovod}.

Circuit cutting provides a principled mechanism to execute circuits that do not fit a target backend by decomposing them into smaller subcircuits that can be executed independently, followed by classical reconstruction of the desired quantity~\cite{Peng2020ClusterSim,Tang2021CutQC,Piveteau2022Knitting,Harrow2025OptimalCuts}. This transforms a single circuit evaluation into a staged distributed workload: a fan-out of subcircuit executions, followed by a global reduction step that reconstructs the target expectation value (and, in learning settings, supports gradient estimation and optimiser updates). The methodological benefits of cutting are clear. The systems implications are less well characterised. Existing work typically reports overhead in terms of subcircuit counts, sampling complexity, or asymptotic reconstruction cost~\cite{Tang2021CutQC,Harrow2025OptimalCuts,Lowe2023FastCutting}, which is informative for feasibility but does not directly answer the operational question that governs scaling: \emph{where does wall-clock time go when we actually run the pipeline in parallel?}

\subsection{Motivation and Problem Statement}

We identify three practical limitations that motivate our study.

\textit{P1: Overhead attribution is under-specified.} Cutting introduces compilation, materialisation, dispatch, and reconstruction costs that are often treated as secondary or summarised coarsely. Without stage-level instrumentation, it is difficult to determine whether slowdowns are intrinsic to cutting or artefacts of the runtime.

\textit{P2: Reconstruction can dominate the critical path.} Cutting increases execution parallelism but concentrates work in a reconstruction stage. Analogous reduction phases bound scalability in distributed computing even when upstream work parallelises well~\cite{DeanGhemawat2004MapReduce,Sergeev2018Horovod}.

\textit{P3: Training dynamics, accuracy, and robustness must be assessed end-to-end.} In learning pipelines, the optimiser observes finite-sample estimators with potentially different noise profiles; small changes can alter optimisation trajectories~\cite{Cerezo2021VQA,Schuld2019Gradients,Endo2021QEM}. Both predictive accuracy and robustness~\cite{Madry2018Robust,Tramer2018EnsembleAdv} must be verified under the same training protocol.

These are systems problems: they concern staging, scheduling, aggregation, and the interaction between distributed execution and iterative optimisation.

\subsection{Research Questions}

We structure the paper around the following research questions (RQs), designed to be answered by measured runtime logs and training traces rather than by asymptotic arguments:

\begin{enumerate}
    \item \textit{RQ1 (Cutting overhead):} What additional end-to-end overheads are introduced by circuit cutting compared to an uncut execution baseline, and how do these overheads scale with the degree of parallelism?
    \item \textit{RQ2 (Scalability bottlenecks):} Which pipeline stages bound scaling under parallel execution, and under what conditions does classical reconstruction become the dominant contributor to wall-clock time?
    \item \textit{RQ3 (Scheduling policy and staggering):} How do execution policies that stagger or otherwise reshape subcircuit dispatch affect throughput, straggler sensitivity, and overall time-to-solution?
    \item \textit{RQ4 (Accuracy and training behaviour):} Does cutting preserve task accuracy under identical training protocols, and can the induced estimator and scheduling characteristics measurably influence training stability (e.g., convergence behaviour) relative to the uncut baseline?
    \item \textit{RQ5 (Robustness):} How does circuit cutting affect robustness metrics under the same evaluation regime, and are observed changes attributable to estimator effects rather than to changes in model definition?
\end{enumerate}

Each RQ is framed to admit an unambiguous operationalisation: we measure stage-level durations, derive scaling trends and bottleneck shares, and evaluate accuracy and robustness from recorded training and evaluation traces.

\subsection{Approach Overview}

To answer these questions, we design an estimator-based execution pipeline that makes cutting explicit as a staged distributed workload: (i)~generate a cut plan and the associated subcircuits~\cite{Tang2021CutQC,Harrow2025OptimalCuts}, (ii)~execute subcircuits as independent tasks on a worker pool, (iii)~reconstruct target estimates via a dedicated aggregation phase, and (iv)~feed reconstructed estimates into an iterative training loop. Each stage is instrumented to produce time-stamped records, enabling stage-level attribution under varying parallelism and scheduling policies. We do not assume proportional speed-ups; instead, we quantify where the critical path shifts as we scale out, and couple runtime measurements with training logs to ensure performance improvements do not come at the cost of model quality (RQ4--RQ5).

The pipeline integrates with the Qiskit machine learning stack: \textit{EstimatorQNN}~\cite{qiskit_machine_learning} wraps the parameterised circuit into a differentiable layer, and \textit{TorchConnector} bridges it to PyTorch. When cutting is enabled, \distest{} intercepts each estimator query, decomposes it into subexperiments via \textit{qiskit-addon-cutting}~\cite{qiskit_addon_cutting}, dispatches them, reconstructs the expectation value, and returns it transparently. Gradients computed via the parameter-shift rule~\cite{Schuld2019Gradients} generate additional estimator queries, each independently decomposed and reconstructed through the same pipeline, ensuring the training loop remains unchanged regardless of whether cutting is enabled.

\subsection{Contributions}

Our main contributions are as follows:

\begin{itemize}
    \item \textit{A measurement-driven formulation of circuit cutting as a distributed staged pipeline}, with instrumentation that attributes end-to-end time across cutting, subcircuit execution, reconstruction, and training, and with explicit integration into the QML training loop via EstimatorQNN and TorchConnector.
    \item \textit{An empirical scalability analysis grounded in stage-level logs}, enabling bottleneck identification and explaining when additional parallelism does not translate into reduced wall-clock time due to aggregation and coordination effects.
    \item \textit{A study of scheduling and straggler sensitivity for cutting workloads}, with a framework supporting eager, batched, and staggered dispatch, evaluated through measured throughput and tail behaviour under straggler injection rather than assumed ideal parallelism.
    \item \textit{An end-to-end assessment of learning outcomes}, reporting accuracy and robustness under consistent training protocols to distinguish performance gains from changes in optimisation behaviour.
    \item \textit{A characterisation of computational scalability barriers}, documenting the exponential growth of reconstruction overhead with cut count and establishing the computational constraints that limit practical experimentation to small qubit counts with current methods.
\end{itemize}

Together, these contributions establish a systems-level understanding of performance trade-offs from circuit cutting in estimator-heavy training pipelines and provide a reproducible basis for reasoning about overheads, bottlenecks, and outcome preservation.

\revised{The rest of the paper is organised as follows. Section~\ref{sec:background} presents the technical background. Section~\ref{sec:related-work} reviews related work and positions our contribution. Section~\ref{sec:system-model} formalises the system model and problem definition. Section~\ref{sec:methodology} describes the methodology and instrumentation. Section~\ref{sec:evaluation} presents the experimental evaluation. Section~\ref{sec:validation} reports a validation on real quantum hardware. Section~\ref{sec:discussion} discusses the findings and future directions. Section~\ref{sec:conclusion} concludes the paper.}

%% file: section-background.tex

\section{Background}
\label{sec:background}

This section summarises the technical background required to interpret our system design and measurements.

\subsection{Hybrid Quantum--Classical Workloads and Variational Training}

Near-term quantum devices operate in the \emph{Noisy Intermediate-Scale Quantum} (NISQ) regime, where limited qubit counts and noise constrain executable circuit width and depth~\cite{Preskill2018NISQ}. A prevalent approach in this regime is the \emph{variational quantum algorithm} (VQA), which couples a parameterised quantum circuit with a classical optimiser~\cite{Peruzzo2014VQE,Cerezo2021VQA,Moll2018QST,Kandala2017HardwareEffVQE}. This hybrid loop underpins much of quantum machine learning (QML), including parameterised quantum models~\cite{Biamonte2017QML,Mitarai2018QCL} and kernel-style feature maps~\cite{Havlicek2019QFeature}. For learning tasks with classical data, practical feasibility often depends as much on data access patterns and estimator variance as on the expressivity of the quantum model~\cite{Huang2021PowerOfData}.

From a systems perspective, VQA training is dominated by repeated evaluation of expectation values and (often) gradients. Analytic gradient schemes such as the parameter-shift rule are attractive because they avoid finite-difference instability but still multiply circuit evaluations~\cite{Schuld2019Gradients}. In addition, the optimisation landscape can be problematic (e.g., barren plateaus), which can increase the number of training iterations and amplify any per-iteration overheads~\cite{McClean2018Barren,Cerezo2021VQA}. Consequently, end-to-end runtime is governed by a composition of quantum execution latency, the number of circuit evaluations required by the optimiser, and classical orchestration and post-processing costs~\cite{Preskill2018NISQ,Cerezo2021VQA}.

Noise and sampling error further complicate this picture. Error mitigation can improve estimator accuracy without full fault tolerance, but it typically adds additional circuit executions and classical processing steps~\cite{Temme2017ErrorMitigation,Endo2021QEM}. The resulting workload is therefore \emph{estimator-heavy}: it comprises large numbers of small quantum jobs plus non-trivial classical aggregation, often executed in an interactive loop~\cite{Endo2021QEM,Cerezo2021VQA}.

\subsection{Estimator Abstractions and Execution Models}

Near-term applications consume quantum hardware through an expectation-value interface: a parameterised circuit and observable(s) yield an estimate with finite-sample variance. Modern stacks formalise this as first-class execution units~\cite{JavadiAbhari2024Qiskit,QiskitPrimitives,Bergholm2018PennyLane,Broughton2020TFQ}. These abstractions make performance failure modes predictable: circuit evaluations grow with estimator variance and optimiser query complexity, aggregation is unavoidable, and the critical path may be dominated by orchestration and reduction rather than raw quantum execution~\cite{Schuld2019Gradients,Temme2017ErrorMitigation,Endo2021QEM}. We explicitly treat this as a distributed systems problem, measuring and attributing overheads across pipeline stages.

\subsection{Circuit Cutting and Classical Reconstruction}

Circuit cutting (often discussed under the umbrella of \emph{circuit knitting}) decomposes a circuit that does not fit a target backend into a collection of smaller subcircuits that do, at the cost of classical reconstruction and increased sampling complexity~\cite{Peng2020ClusterSim,Tang2021CutQC,Piveteau2022Knitting}. The core trade-off is structural: cutting reduces spatial requirements (qubits) while increasing the number of executions and the weight of post-processing~\cite{Peng2020ClusterSim,Tang2021CutQC}.

\begin{figure}[t]
  \centering
  \includegraphics[width=\columnwidth]{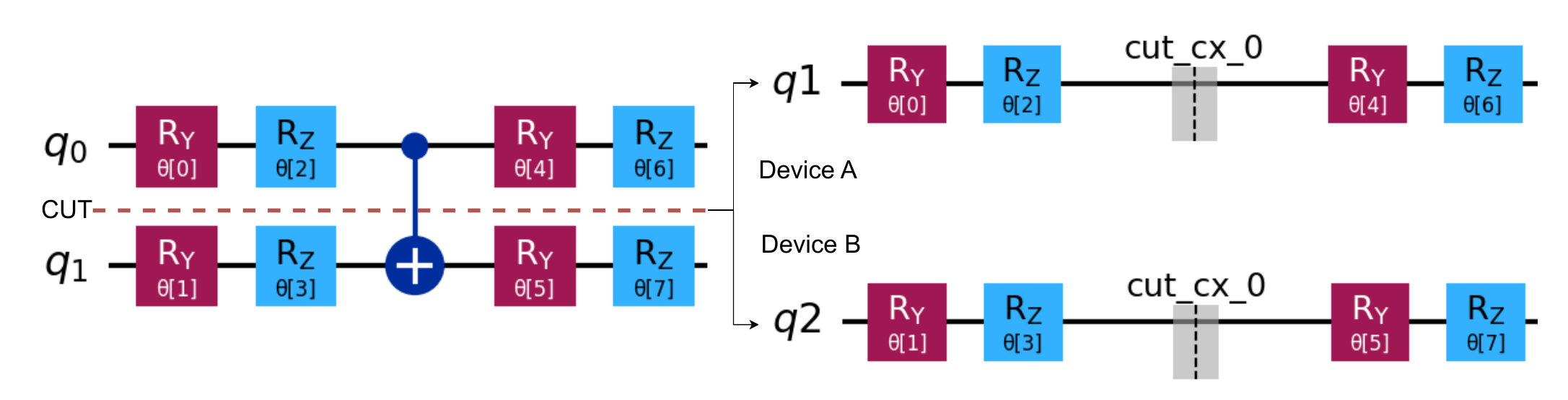}
  \caption{Independent subcircuits generated by cutting a circuit. The dashed line indicates the cut location; the resulting fragments (left and right subcircuits) can be executed independently on smaller devices. Each cut exponentially increases the number of subexperiments and reconstruction terms.}
  \label{fig:cut}
\end{figure}

Practical toolflows typically involve three stages. First, a compiler or optimiser chooses cut locations and a decomposition strategy~\cite{Tang2021CutQC,Harrow2025OptimalCuts}. Second, a set of subcircuits (often many variants induced by the decomposition) are executed and measured~\cite{Tang2021CutQC,Lowe2023FastCutting}. Third, a classical reconstruction procedure aggregates sub-results into an estimate of the target quantity (e.g., an expectation value)~\cite{Tang2021CutQC,Piveteau2022Knitting,Harrow2025OptimalCuts}. Recent work shows that allowing classical communication between subcomponents can reduce overhead constants in certain settings~\cite{Piveteau2022Knitting}, and that alternative cutting schemes based on randomised measurements provide different accuracy--cost trade-offs~\cite{Lowe2023FastCutting}.

Importantly for systems evaluation, cutting turns a single ``large'' circuit evaluation into a \emph{task graph} with (often) substantial fan-out followed by a global aggregation. This introduces obvious bottleneck candidates: scheduling overhead, straggler sensitivity across large job sets, and a potentially heavy reduction (reconstruction) stage~\cite{Tang2021CutQC,Harrow2025OptimalCuts}. These are precisely the failure modes we can only resolve by measured profiling rather than by asymptotic overhead counts alone.

The overhead scaling of circuit cutting is exponential in the number of cuts: each additional cut multiplies the number of required subexperiments and reconstruction terms. For $c$ cuts across CNOT gates, the number of subexperiments grows as $\mathcal{O}(9^c)$ in the wire-cutting decomposition employed by \textit{qiskit-addon-cutting}~\cite{qiskit_addon_cutting,Tang2021CutQC}, and the classical reconstruction cost scales correspondingly. This exponential growth has profound practical consequences that we quantify empirically in our evaluation (Section~\ref{sec:evaluation}).

\revised{%
\paragraph{How reconstruction works and why it is costly.}
Because reconstruction is the stage we identify as the dominant bottleneck, we describe its mechanics explicitly. Cutting replaces each non-local operation at a cut by a \emph{quasiprobability decomposition} into locally realisable operations: a two-qubit channel spanning a cut is written as a weighted sum $\sum_{a} c_a\, (A_a \otimes B_a)$, where $A_a$ and $B_a$ act only within their respective fragments and the real coefficients $c_a$ may be \emph{negative}~\cite{Piveteau2022Knitting,Tang2021CutQC}. For a wire cut realised through CNOT decomposition this expansion contributes a fixed number of basis terms per cut, with the well-known sampling overhead $\gamma=9$ entering through $\sum_a |c_a|$. After the induced subcircuit variants are executed and the local expectation value of each subobservable is estimated on each fragment $f$, the target expectation value is reconstructed as a coefficient-weighted sum over the \emph{tensor products} of the per-fragment outcomes,
\begin{equation}
\langle O \rangle \;\approx\; \sum_{\mathbf{a}} \Big(\textstyle\prod_{i \in \text{cuts}} c_{a_i}\Big)\, \prod_{f \in \text{fragments}} \langle o_f \rangle_{\mathbf{a}},
\label{eq:reconstruction}
\end{equation}
where the outer sum runs over all joint assignments $\mathbf{a}=(a_1,\dots,a_c)$ of decomposition terms to the $c$ cut locations. The cost of reconstruction materialises from two sources. First, the number of joint assignments---and hence of coefficient-weighted product terms that must be formed and summed---grows multiplicatively with each cut, reaching $\mathcal{O}(9^c)$, because each cut independently multiplies the number of basis terms to be combined. Second, for every term the procedure performs a tensor-product combination of fragment results together with bookkeeping over the subobservable and measurement-basis combinations per fragment. Crucially, this is \emph{purely classical} post-processing whose cost is intrinsic to the decomposition and, unlike quantum execution time, does not diminish with improved quantum hardware. This is the structural reason reconstruction can come to dominate the per-query critical path as the number of cuts increases.}

\subsection{Distributed Execution, Stragglers, and Staged Pipelines}

Distributed systems research has shown that scalability is frequently capped by coordination and stragglers rather than by idealised parallel work~\cite{DeanGhemawat2004MapReduce,Zaharia2008LATE}. Speculative execution (e.g., LATE~\cite{Zaharia2008LATE}), in-memory dataflow~\cite{Zaharia2012RDD}, and task-graph engines such as Ray~\cite{Moritz2018Ray} address these bottlenecks at different levels. In distributed ML, parameter-server designs~\cite{Li2014ParameterServer,Ho2013SSP}, lock-free updates~\cite{Recht2011Hogwild}, and collective communication~\cite{Sergeev2018Horovod,Cho2019BlueConnect} all show that aggregation phases can dominate at scale even when compute parallelises cleanly~\cite{Goyal2017LargeBatch,BlumofeLeiserson1999WorkStealing}. Circuit cutting produces a staged pipeline with the same fundamental shape---embarrassingly parallel subcircuit execution followed by a global reconstruction barrier---so the relevant question is whether reconstruction and coordination costs dominate the critical path as we scale out~\cite{Tang2021CutQC,Zaharia2008LATE,Sergeev2018Horovod}. Our evaluation quantifies this boundary empirically.

%% file: section-related-work.tex

\section{Related Work}
\label{sec:related-work}

We organise related work into (i)~circuit cutting methods and their overhead models, (ii)~hybrid software stacks and execution interfaces, and (iii)~distributed systems techniques for scaling estimator-heavy workloads.

\subsection{Circuit Cutting Methods, Overheads, and Optimisation Targets}

Peng et al.~\cite{Peng2020ClusterSim} introduced a cluster simulation view motivating decomposition as a route to execute circuits beyond device width. CutQC~\cite{Tang2021CutQC} provided an end-to-end toolflow documenting that post-processing and subcircuit counts grow rapidly with cuts. Subsequent work reduced overhead via classical communication~\cite{Piveteau2022Knitting} and randomised measurements~\cite{Lowe2023FastCutting}, while Harrow et al.~\cite{Harrow2025OptimalCuts} formalised optimality notions and bounds. Complementary work on multi-processor execution~\cite{CarreraVazquez2024MultiQPU} and serverless cutting environments~\cite{Garrison2023ServerlessCutting,Sitdikov2023CKT,Johnson2023SCPoster} acknowledges that orchestration and classical processing are first-order concerns. However, the literature reports overhead primarily in \emph{counts} rather than \emph{measured} end-to-end time attribution, limiting understanding of when reconstruction becomes the bottleneck and whether scheduling policies materially affect throughput.

\subsection{Hybrid Stacks and Estimator-Oriented Interfaces}

The hybrid execution ecosystem has matured: Qiskit~\cite{JavadiAbhari2024Qiskit} and its primitives~\cite{QiskitPrimitives} make expectation estimation explicit, while PennyLane~\cite{Bergholm2018PennyLane} and TensorFlow Quantum~\cite{Broughton2020TFQ} integrate quantum evaluation into differentiable ML workflows. These stacks generally do not analyse distributed bottlenecks for cutting-plus-training pipelines. Error mitigation adds further circuit evaluations and classical processing~\cite{Temme2017ErrorMitigation,Endo2021QEM}, and measurement cost is a central constraint~\cite{Cerezo2021VQA,Moll2018QST}. \revised{A related line of research is \emph{quantum architecture search} (QAS), which automates the design of the parameterised circuit itself. Recent work uses curriculum reinforcement learning to discover architectures that remain performant under realistic hardware error profiles~\cite{Patel2024CurriculumRLQAS}, explicitly targeting the gap between ideal circuit expressivity and device-imposed constraints---the same gap that circuit cutting addresses from the execution side. QAS is orthogonal and complementary to our contribution: it selects \emph{which} circuit to run, whereas we study the distributed systems cost of \emph{executing} a fixed circuit under cutting. A QAS-discovered architecture that exceeds device width would still rely on cutting to run, and would therefore inherit exactly the reconstruction and coordination bottlenecks we characterise; conversely, the per-query overhead profile we measure could serve as a hardware-aware cost term inside a QAS objective.} Yet, the interaction between these algorithmic requirements and distributed execution policies is typically left implicit. Our work treats this interaction as the primary object of measurement.

\subsection{Distributed ML and Straggler-Aware Scheduling}

Distributed ML provides directly relevant methods: parameter-server and bounded-staleness models manage heterogeneity~\cite{Li2014ParameterServer,Ho2013SSP}, speculative execution mitigates tail latency~\cite{Zaharia2008LATE}, and collective aggregation can dominate at scale~\cite{Sergeev2018Horovod,Cho2019BlueConnect,Goyal2017LargeBatch}. Classical adversarial robustness work~\cite{Szegedy2013Intriguing,Goodfellow2015Adversarial,Madry2018Robust,Tramer2018EnsembleAdv} shows that training dynamics materially affect robustness, and analogous concerns arise in estimator-heavy quantum training from sampling noise and distribution shifts induced by decomposition~\cite{Cerezo2021VQA,Schuld2019Gradients,Endo2021QEM}. Existing work rarely connects these robustness questions to the systems bottlenecks of cutting and reconstruction.

\subsection{Comparison with Existing Approaches}

Table~\ref{tab:comparison} summarises how our work relates to the most directly comparable circuit-cutting studies across five dimensions relevant to practical deployment.

\begin{table*}[t]
  \centering
  \caption{Comparison with key related work on circuit cutting. \textbf{Approach}: theoretical analysis (T), end-to-end toolflow (TF), or systems profiling (SP). \textbf{Overhead analysis}: asymptotic bounds (A), empirical measurement (E), or both. \textbf{Training integration}: whether cutting is embedded in an iterative variational training loop. \textbf{Hardware evaluation}: whether results on a real QPU are reported. \textbf{Bottleneck identification}: whether specific pipeline stages are measured and identified as scaling limiters.}
  \label{tab:comparison}
  \small
  \setlength{\tabcolsep}{5pt}
  \begin{tabular}{llcccc}
    \toprule
    \textbf{Work} & \textbf{Approach} & \textbf{Overhead} & \textbf{Training} & \textbf{Hardware} & \textbf{Bottleneck} \\
    & & \textbf{analysis} & \textbf{integration} & \textbf{evaluation} & \textbf{identification} \\
    \midrule
    Peng et al.~\cite{Peng2020ClusterSim} & T & A & $\times$ & $\times$ & $\times$ \\
    Tang et al.\ (CutQC)~\cite{Tang2021CutQC} & TF & A+E & $\times$ & \checkmark & $\times$ \\
    Piveteau \& Sutter~\cite{Piveteau2022Knitting} & T & A & $\times$ & $\times$ & $\times$ \\
    Lowe et al.~\cite{Lowe2023FastCutting} & T & A & $\times$ & $\times$ & $\times$ \\
    Harrow \& Lowe~\cite{Harrow2025OptimalCuts} & T & A & $\times$ & $\times$ & $\times$ \\
    Carrera V.\ et al.~\cite{CarreraVazquez2024MultiQPU} & TF & E & $\times$ & \checkmark & $\times$ \\
    \textbf{Our work} & \textbf{SP} & \textbf{E} & \checkmark & \checkmark & \checkmark \\
    \bottomrule
  \end{tabular}
\end{table*}

\subsection{Summary and Gap}

Prior work establishes that circuit cutting extends executable circuit sizes and clarifies overhead trade-offs~\cite{Peng2020ClusterSim,Tang2021CutQC,Piveteau2022Knitting,Lowe2023FastCutting,Harrow2025OptimalCuts}, while hybrid stacks expose estimator-oriented execution~\cite{JavadiAbhari2024Qiskit,QiskitPrimitives,Bergholm2018PennyLane,Broughton2020TFQ} and distributed systems research provides mature straggler and aggregation techniques~\cite{DeanGhemawat2004MapReduce,Zaharia2008LATE,Li2014ParameterServer,Sergeev2018Horovod}. What remains underexplored is a measurement-driven account of how cutting transforms end-to-end training into a staged distributed pipeline and which execution policies materially improve throughput without degrading accuracy or robustness. We address this gap by instrumenting costs across cutting, execution, reconstruction, and training dynamics.

%% file: section-system-model.tex

\section{System Model and Problem Formulation}
\label{sec:system-model}

In this section, we formalise the estimator-driven, cut-aware training pipeline as a staged distributed computation and define measurable performance and outcome quantities aligned with our research questions (RQ1--RQ5).

\subsection{End-to-End Pipeline Model}

\begin{figure}[t]
  \centering
  \includegraphics[width=\columnwidth]{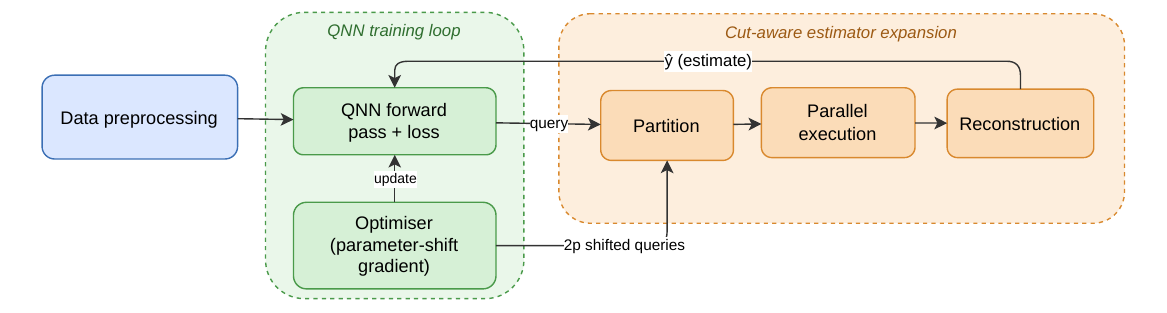}
  \caption{\revised{Training pipeline with a cut-aware distributed estimator, laid out left-to-right. \textbf{Left (blue):} classical data preprocessing. \textbf{Centre (green):} the QNN training loop, comprising the forward pass/loss and the optimiser. \textbf{Right (orange):} the cut-aware estimator expansion---partitioning, parallel execution, and classical reconstruction. Crucially, the expansion is triggered by \emph{both} the forward pass \emph{and} the optimiser: the parameter-shift rule issues $2p$ shifted estimator queries per iteration (for $p$ parameters), each of which independently undergoes the same partition--execute--reconstruct expansion. The reconstructed estimate $\widehat{y}$ is returned to the training loop.}}
  \label{fig:pipeline}
\end{figure}

Figure~\ref{fig:pipeline} shows the pipeline we evaluate. A classical dataset is preprocessed and supplied to a training loop. Each iteration evaluates a quantum neural network (QNN) forward pass and associated loss, and updates parameters via a classical optimiser. The forward pass is implemented via expectation estimation queries of the form $(C(\theta,x), O)$, where $C$ is a parameterised circuit instantiated at parameters $\theta$ and input $x$, and $O$ is the observable (or set of observables) required to compute model outputs and loss~\cite{Cerezo2021VQA,Schuld2019Gradients}.

When circuit cutting is enabled, the estimator query is expanded into a set of subexperiments that can be executed independently, followed by a classical reconstruction step that returns the estimate to the training loop~\cite{Tang2021CutQC,Peng2020ClusterSim,Harrow2025OptimalCuts}. This transformation is central to our study because it introduces (i)~additional overheads beyond baseline execution (RQ1), (ii)~a reconstruction barrier that can dominate the critical path under scale-out (RQ2), and (iii)~scheduling degrees of freedom, including staggering, that affect straggler sensitivity and throughput (RQ3). The same transformation may also affect learning dynamics through estimator variance and evaluation order, which we examine through accuracy and robustness outcomes (RQ4--RQ5).

\subsection{Integration with QML Training Loop}
\label{subsec:qml-integration}

To make the integration between the cut-aware pipeline and the QML training loop precise, we describe the data and control flow through the key software components.

The parameterised quantum circuit $C(\theta, x)$ and observable $O$ are encapsulated in an \textit{EstimatorQNN} instance~\cite{qiskit_machine_learning}, which exposes a differentiable forward pass and gradient computation interface. The \textit{TorchConnector} wraps this \textit{EstimatorQNN} as a PyTorch \texttt{nn.Module}, enabling standard PyTorch optimisers (e.g., Adam) and loss functions (e.g., cross-entropy) to be used without modification.

During the forward pass, \textit{TorchConnector} calls \textit{EstimatorQNN.forward()}, issuing estimator queries $(C(\theta, x_i), O)$ for each sample $x_i$. In the baseline configuration, queries are evaluated directly by the Aer simulator; when cutting is enabled, \distest{} intercepts each query and processes it through the staged pipeline below.

For gradient computation, the \textit{EstimatorQNN} uses the parameter-shift rule~\cite{Schuld2019Gradients}: two additional queries with $\theta_j \pm \pi/2$ per parameter $\theta_j$, each independently processed through the cut-aware pipeline. The cut-aware estimator is transparent to the optimiser: gradients are mathematically equivalent to the uncut case up to finite-sample variance. The overhead is multiplicative: the total cut-aware estimator invocations per iteration is $(q + 2p) \times |\text{batch}|$ (for $q$ forward queries and $p$ parameters), each expanding into $K$ subexperiments---the root cause of the training time overheads documented in RQ1.

\subsection{Cut-Aware Estimator Query Model}

For a single query instance $(C(\theta,x), O)$, circuit cutting produces a partition labelled by $\ell$ and yields subcircuits and subobservables. The resulting execution requires a finite set of concrete subexperiments $\mathcal{E}=\{e_k\}_{k=1}^{K}$ and reconstruction coefficients $\alpha$ determined by the cutting scheme and partition~\cite{Tang2021CutQC,Harrow2025OptimalCuts}. Each $e_k$ is executed with a fixed shot count $S$, producing a classical summary $r_k$ (e.g., outcome counts). Reconstruction aggregates $\mathcal{R}=\{r_k\}$ into an estimate $\widehat{y}$ of the target expectation value.

This induces a staged cost decomposition for each query instance:
\begin{align}
T_{\mathrm{total}}
= T_{\mathrm{part}} + T_{\mathrm{gen}} + T_{\mathrm{exec}} + T_{\mathrm{rec}},
\label{eq:ttotal}
\end{align}
where:
\begin{itemize}
  \item $T_{\mathrm{part}}$ is circuit/observable partitioning time (including cut planning and partition materialisation);
  \item $T_{\mathrm{gen}}$ is subexperiment generation time (including basis variants and reconstruction coefficients);
  \item $T_{\mathrm{exec}}$ is parallel execution time of $\mathcal{E}$ on a worker pool;
  \item $T_{\mathrm{rec}}$ is classical reconstruction time.
\end{itemize}
Eq.~\eqref{eq:ttotal} provides the measurement backbone for RQ1 and RQ2: we report each component as a measured quantity and study how their shares evolve with the degree of parallelism.

\revised{%
\textbf{Amortisation of structural stages across training.}
Eq.~\eqref{eq:ttotal} decomposes the cost of a \emph{single} query, but in an iterative training loop the circuit \emph{structure} $C(\cdot,\cdot)$---its gates, wires, and cut locations---is invariant across queries; only the parameter values $(\theta,x)$ change between iterations. Consequently, partitioning and the generation of subexperiment \emph{templates} (together with the reconstruction coefficients $\alpha$, which depend only on the cut structure) need to be computed \emph{once} for a given partition label and then reused. Our implementation caches these structural artefacts keyed on the circuit structure and partition label, so that for every subsequent query partitioning and generation are not repeated: each query instead pays only a lightweight parameter-binding cost $T_{\mathrm{bind}}$ that rebinds $(\theta,x)$ into the cached templates. The per-query cost over a training run of $N$ queries is therefore more precisely
\begin{align}
T_{\mathrm{total}}^{(j)} =
\underbrace{\big(T_{\mathrm{part}}+T_{\mathrm{gen}}\big)\,\mathbf{1}\{j=1\}}_{\text{one-time, cached}}
+ T_{\mathrm{bind}}^{(j)} + T_{\mathrm{exec}}^{(j)} + T_{\mathrm{rec}}^{(j)},
\label{eq:ttotal-amortised}
\end{align}
for the $j$-th query ($j=1,\dots,N$), so that the amortised contribution of $T_{\mathrm{part}}+T_{\mathrm{gen}}$ vanishes as $1/N$ over a training run. This corrects a redundancy that would otherwise artificially inflate the per-iteration overhead by re-deriving structural quantities at every step. For the small circuits studied here, $T_{\mathrm{part}}$ and $T_{\mathrm{gen}}$ are already small relative to $T_{\mathrm{exec}}+T_{\mathrm{rec}}$, so caching leaves the reported stage \emph{shares} essentially unchanged; for larger and deeper circuits, where cut planning and template generation become expensive (Section~\ref{sec:evaluation}), this amortisation is essential and shifts the relevant per-iteration cost onto $T_{\mathrm{bind}}$, $T_{\mathrm{exec}}$, and $T_{\mathrm{rec}}$.}

\subsection{Execution, Stragglers, and Scheduling}

\begin{figure}[t]
  \centering
  \includegraphics[width=\columnwidth]{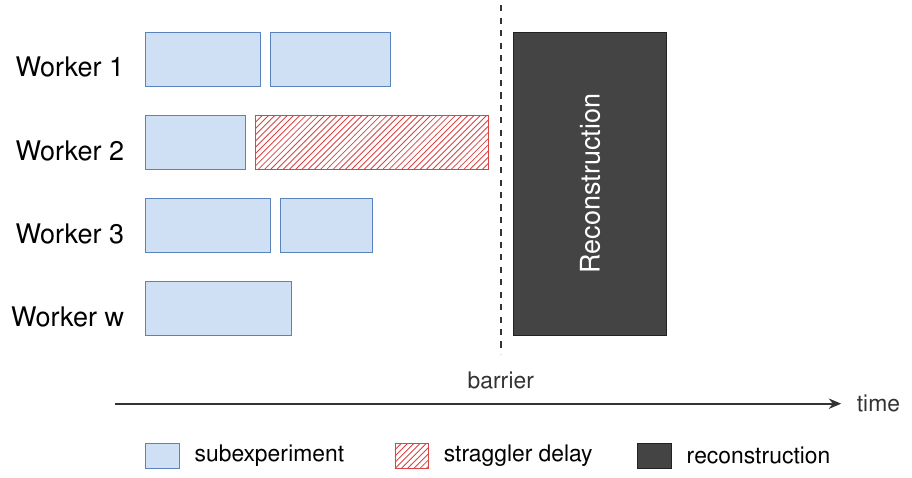}
  \caption{\revised{Per-query expansion under circuit cutting. Subexperiments (light shaded blocks) execute across $w$ workers; classical reconstruction (dark block) forms a barrier that cannot begin until all subexperiment results are available. A straggler delay (hatched block, Worker~2) extends that worker's completion time and therefore pushes back the barrier, delaying reconstruction and lengthening the critical path.}}
  \label{fig:scheduling}
\end{figure}

We model subexperiment execution as a set of independent tasks scheduled on $w$ workers, followed by a reconstruction barrier. Figure~\ref{fig:scheduling} captures this per-query expansion and the straggler sensitivity introduced by the barrier. Let $t_k$ denote the service time for subexperiment $e_k$ (including dispatch and backend runtime). Under a work-conserving policy, the realised parallel time is governed by load balance and tail effects and can be expressed as:
\begin{align}
T_{\mathrm{exec}} \approx \max_{i \in \{1,\dots,w\}} \sum_{k \in \mathcal{A}(i)} t_k,
\label{eq:texec}
\end{align}
where $\mathcal{A}(i)$ is the assignment of subexperiments to worker $i$. This form makes explicit why speed-up can saturate even when the task set is abundant: $T_{\mathrm{exec}}$ is bounded by the slowest worker, and $T_{\mathrm{rec}}$ is a serial reduction that must wait for all results, mirroring classical bottlenecks in staged distributed computations~\cite{DeanGhemawat2004MapReduce,Zaharia2008LATE}.

To study straggler sensitivity repeatably, we allow an optional synthetic delay process at the task level. Each subexperiment independently incurs an injected delay $\Delta$ with probability $p$, yielding $t_k' = t_k + \mathbf{1}\{u_k<p\}\Delta$ for $u_k\sim\mathrm{Uniform}(0,1)$. We use this mechanism strictly as an experimental control to stress the tail of Eq.~\eqref{eq:texec} and to evaluate whether scheduling choices (including staggering) reduce end-to-end time-to-solution under heterogeneity~\cite{Zaharia2008LATE}.

\subsection{Problem Definition}

We consider a fixed training protocol (dataset split, optimiser, stopping criterion) and a fixed per-subexperiment shot budget~$S$. For a given configuration (cut plan~$\ell$, worker pool size~$w$, scheduling policy~$\pi$, and optional straggler parameters), our primary systems objective is to minimise \emph{time-to-solution} while preserving learning outcomes:
\begin{align}
\min_{\pi \in \Pi} \ \mathbb{E}[T_{\mathrm{total}}]
\quad \text{s.t.} \quad
\text{accuracy and robustness constraints.}
\end{align}
We operationalise this objective through:
\begin{itemize}
  \item \textit{Measured:} $T_{\mathrm{part}}, T_{\mathrm{gen}}, T_{\mathrm{exec}}, T_{\mathrm{rec}}, T_{\mathrm{total}}$ per query instance (and their aggregation over training);
  \item \textit{Derived:} stage shares (e.g., $T_{\mathrm{rec}}/T_{\mathrm{total}}$), scaling curves as a function of $w$, and bottleneck attribution via changes in dominant stage across configurations;
  \item \textit{Observed:} training outcomes (accuracy) and robustness metrics computed under a consistent evaluation regime.
\end{itemize}
This mapping supports RQ1--RQ5: overhead and scaling from timing breakdowns, reconstruction bottleneck from measured $T_{\mathrm{rec}}$ dominance and speed-up saturation, staggering impact from $T_{\mathrm{exec}}$ tail behaviour changes, and outcome preservation from accuracy and robustness under unchanged evaluation procedures.

%% file: section-methodology.tex

\section{Methodology}
\label{sec:methodology}

In this section, we describe how we implement the cut-aware distributed estimator, how we instrument stage-level timings and outcomes, and how we configure scheduling policies and straggler controls to answer RQ1--RQ5.

\subsection{Cut-Aware Distributed Estimator}

Our estimator backend implements a cut-aware execution path that transforms each query instance $(C(\theta,x), O)$ into a set of subexperiments and a reconstruction plan. Concretely, for each query instance we:
(i)~partition the problem according to a partition label $\ell$,
(ii)~generate the concrete subexperiments $\mathcal{E}$ and reconstruction coefficients $\alpha$,
(iii)~execute $\mathcal{E}$ across a worker pool of size $w$,
(iv)~reconstruct the target estimate $\widehat{y}$,
(v)~return $\widehat{y}$ to the training loop.

The key methodological choice is to treat this as a \emph{staged} pipeline and to instrument each stage individually. This allows us to attribute overheads introduced by cutting (RQ1) and to identify scaling limits caused by reconstruction and coordination (RQ2) without conflating them with training-loop compute.

Algorithm~\ref{alg:estimator} summarises the per-query logic. The algorithm is intentionally phrased to align with logged quantities and the timing decomposition in Eq.~\eqref{eq:ttotal}.

\begin{algorithm}[t]
\caption{Instrumented cut-aware estimator for one query instance}
\label{alg:estimator}
\begin{algorithmic}[1]
\REQUIRE Circuit $C(\theta,x)$, observable $O$, partition label $\ell$, \revised{QPD sample budget $M$,} measurement shots $S$, workers $w$, scheduling policy $\pi$
\STATE Initialise timers and metadata record
\IF{\revised{cut plan for $(C\text{-structure},\ell)$ not cached}}
  \STATE \textbf{Partition:}
  $(\{C_i\}, \{O_i\}, \mathcal{B}) \leftarrow \mathrm{PartitionProblem}(C,O,\ell)$
  \STATE \textbf{Generate:}
  $(\mathcal{E}, \alpha) \leftarrow \mathrm{GenerateSubexperiments}(\{C_i\},\{O_i\},\revised{M})$
  \STATE \revised{Cache $(\mathcal{E},\alpha,\{O_i\})$ keyed on $(C\text{-structure},\ell)$}
\ENDIF
\STATE \revised{\textbf{Bind:} $\mathcal{E}_{\theta,x} \leftarrow \mathrm{BindParameters}(\mathcal{E}, \theta, x)$}
\STATE \textbf{Execute:} $\mathcal{R} \leftarrow \mathrm{ExecuteTasks}(\mathcal{E}\revised{_{\theta,x}}, \revised{S,} w, \pi)$
\STATE \textbf{Reconstruct:} $\widehat{y} \leftarrow \mathrm{Reconstruct}(\mathcal{R}, \alpha, \{O_i\}, O)$
\STATE Log $(T_{\mathrm{part}},T_{\mathrm{gen}},\revised{T_{\mathrm{bind}},}T_{\mathrm{exec}},T_{\mathrm{rec}},T_{\mathrm{total}})$ and metadata to JSONL
\RETURN $\widehat{y}$
\end{algorithmic}
\end{algorithm}

\revised{%
Three aspects of Algorithm~\ref{alg:estimator} warrant clarification. First, $\mathcal{B}=\{\beta_i\}$ is the set of quasiprobability \emph{bases} produced by partitioning: each cut contributes one basis $\beta_i$ with an individual sampling overhead $\gamma_i$, and the total sampling overhead of the cut plan is the product $\prod_i \gamma_i$ (e.g.\ $9^c$ for $c$ CNOT-based wire cuts). $\mathcal{B}$ thus determines the per-query overhead we report and is otherwise consumed by generation. Second, the generation step takes the \emph{QPD sample budget} $M$---the number of quasiprobability samples used to represent the decomposition of the cut channels---which is distinct from the \emph{measurement shots} $S$ applied when each subexperiment is executed on the backend (step~8). The two are commonly conflated: $M$ controls how faithfully the quasiprobability decomposition is represented and may be taken to be exact ($M\!\to\!\infty$), whereas $S$ controls per-circuit sampling variance. Third, because partitioning and generation depend only on the circuit \emph{structure} and partition label---not on $(\theta,x)$---they are computed once and cached (steps~2--6), and every subsequent query reuses the cached templates and pays only the parameter-binding cost of step~7, consistent with the amortised cost model in Eq.~\eqref{eq:ttotal-amortised}.}

\subsection{Scheduling Policies and Staggering}
\label{subsec:scheduling}

We evaluate execution policies that differ in how they submit the task set $\mathcal{E}$ to a bounded worker pool of size $w$. We consider three dispatch strategies: (1)~\textbf{Eager dispatch}, where all subexperiments are submitted immediately under work-conserving, first-come-first-served allocation (our baseline); (2)~\textbf{Batched dispatch}, where subexperiments are partitioned into batches of size $B$ with inter-batch delay $\delta$ to control concurrency; and (3)~\textbf{Staggered dispatch}, where subexperiments are grouped by structural similarity and spaced to avoid correlated queueing delays. All three terminate in the same reconstruction barrier. A policy that reduces task-completion variance can lower $\max_i \sum_{k \in \mathcal{A}(i)} t_k$ and thereby reduce $T_{\mathrm{exec}}$ without changing total work.

Our current evaluation focuses on eager dispatch with straggler injection to stress the tail of $T_{\mathrm{exec}}$ (RQ3). A systematic comparison of batched and staggered policies is an important future direction; the instrumentation framework supports this extension, with all dispatch parameters logged alongside stage-level timings.

\subsection{Straggler Control}

When enabled, synthetic straggler injection is applied at task execution time. Each subexperiment independently incurs an additional delay $\Delta$ with probability $p$, yielding $t_k' = t_k + \mathbf{1}\{u_k < p\}\Delta$. This allows us to stress the barrier sensitivity in Fig.~\ref{fig:scheduling} and evaluate whether the critical-path structure (execution-dominated vs. reconstruction-dominated) modulates straggler impact (RQ3). We log injected delays alongside per-stage timings to ensure that straggler effects are explicitly traceable in subsequent analysis.

\subsection{Instrumentation and Outcome Tracking}

For each estimator query, we emit a JSONL record containing query structure ($K$, $S$), configuration ($\ell$, $w$, scheduling policy, straggler parameters), and stage timings per Eq.~\eqref{eq:ttotal}. Training-level logs capture loss traces, optimiser steps, and evaluation metrics. This supports overhead attribution (RQ1), scaling saturation analysis (RQ2), straggler sensitivity quantification (RQ3), and outcome preservation assessment (RQ4--RQ5) under controlled, matched comparisons.

%% file: section-evaluation.tex

\section{Performance Evaluation}
\label{sec:evaluation}

We evaluate the proposed cut-aware estimator pipeline using instrumented execution traces to quantify end-to-end overheads, scaling behaviour, sensitivity to injected stragglers, and learning outcomes (accuracy and robustness). All claims in this section are grounded in measured quantities recorded in the run-level records, estimator-call summaries, and robustness traces.

\subsection{Experimental Setup}
\label{subsec:setup}

\textbf{Software stack.}
All experiments are implemented in Python. Quantum circuits are constructed in Qiskit~\cite{qiskit} and executed using the Aer primitive interface (SamplerV2/AerSampler, depending on the installed \textit{qiskit-aer} version). Circuit cutting is implemented with \textit{qiskit-addon-cutting}~\cite{qiskit_addon_cutting} via the standard decomposition workflow, which performs wire cutting through CNOT gate decomposition. The learning pipeline uses \textit{qiskit-machine-learning} (EstimatorQNN and TorchConnector~\cite{qiskit_machine_learning}) and PyTorch for optimisation; MNIST~\cite{mnist} data handling uses Torchvision, and Iris~\cite{iris} preprocessing uses scikit-learn. The distributed execution model uses a bounded thread-based worker pool, with the worker count logged as \textit{subexp\_workers} (MNIST) or \textit{max\_subexp\_workers} (Iris).

\revised{%
\textbf{Hardware platform.}
The classical host for all simulator-based experiments and for the classical reconstruction stage is an Apple M4 system-on-chip (10 cores) with 16~GB of unified memory running macOS~26.5. The full software stack comprises Python~3.10, Qiskit~2.3.1, \textit{qiskit-aer}~0.17.2, \textit{qiskit-machine-learning}~0.9.0, \textit{qiskit-ibm-runtime}~0.46.1, and PyTorch~2.11. For the real-hardware validation reported in Section~\ref{sec:validation}, subexperiments are executed on the IBM Quantum \texttt{ibm\_kingston} superconducting processor (an IBM~Heron~r2 device, 156 qubits) accessed through the \textit{qiskit-ibm-runtime} Sampler primitive; circuits are transpiled to the device basis with a preset pass manager at optimisation level~1, and reconstruction is performed on the classical host above. We report the device backend, transpilation settings, and shot budget alongside every hardware measurement so that platform-dependent timing effects are explicitly attributable.}

\textbf{Model circuits.}
Across both workloads, we instantiate a parameterised quantum circuit as the composition of a \textit{ZFeatureMap} (encoding classical features into qubit rotations) followed by a \textit{RealAmplitudes} ansatz with one repetition (\textit{reps}=1), which provides a hardware-efficient parameterisation of the unitary search space. For Iris, the observable is the tensor-product Pauli-$Z$ operator $Z^{\otimes n}$, implemented as a \textit{SparsePauliOp} with $n$ qubits, yielding a scalar expectation-value classifier output. For MNIST, a similar observable structure is used. The choice of a single-repetition ansatz keeps the parameter count manageable while providing sufficient expressivity for the binary classification tasks considered.

\textbf{Workloads and training budgets.}
We consider two binary classification tasks: \textit{iris\_binary\_pm1} (two classes from the Iris dataset, labels mapped to $\pm 1$) and \textit{mnist\_binary} (two-digit binary classification from MNIST). We enforce matched comparisons within each dataset by fixing the estimator shot budget (1024 shots per (sub)experiment) and by restricting comparisons to runs executed under the same training budget and straggler configuration.

For the clean (non-straggler) setting, we report results using the longest training budgets available for each dataset (Iris: $\textit{maxiter}=60$; MNIST: 10 epochs), with \textit{straggler\_delay\_s}$=0.0$. For scaling analysis, shorter budgets are used (Iris: $\textit{maxiter}=10$; MNIST: 5 epochs), again with \textit{straggler\_delay\_s}$=0.0$, to match the worker-count sweeps used in our scaling experiments. For straggler sensitivity, we inject delays of \textit{straggler\_delay\_s}$=0.1$; these experiments use Iris with $\textit{maxiter}=10$ and MNIST with 3 epochs at 8 workers.

\textbf{Computational cost and qubit-count scope.}
\label{subsubsec:qubit-limitation}
The computational cost of training hybrid quantum models with circuit cutting grows exponentially with the number of cuts. In the wire-cutting decomposition used by \textit{qiskit-addon-cutting}, each cut across a CNOT gate decomposes the gate into nine subexperiment variants, yielding a multiplicative overhead of $\mathcal{O}(9^c)$ for $c$ cuts~\cite{qiskit_addon_cutting,Tang2021CutQC}. This means that a circuit with one cut requires $9\times$ more subexperiments than the uncut baseline, two cuts require $81\times$, and three cuts require $729\times$. In addition to the subexperiment count, the classical reconstruction cost and the intermediate state that must be stored and processed grow correspondingly.

\revised{This constraint is a property of \emph{cut-aware training}, not of circuit cutting in isolation. Single-shot circuit-cutting toolflows such as CutQC routinely target much larger circuits---on the order of tens to over a hundred qubits---because they evaluate a circuit \emph{once}~\cite{Tang2021CutQC,Harrow2025OptimalCuts}. Our setting differs fundamentally: we embed cutting inside an \emph{iterative} variational training loop. As established in Section~\ref{subsec:qml-integration}, each training run issues $(q + 2p)\times|\text{batch}|$ estimator queries \emph{per iteration} (the parameter-shift gradient contributes $2p$ queries for $p$ parameters), and \emph{every} such query independently expands by the $\mathcal{O}(9^c)$ factor. The quantity that becomes prohibitive is therefore the \emph{product} of the cutting overhead and the training-query volume, not the cutting overhead alone. It is this product---compounded across iterations, dataset samples, seeds, and worker-count sweeps---that restricts systematic experimentation to small circuit widths, even though a single cut evaluation at these cut counts is itself inexpensive.}

\revised{To separate the \emph{structural} (one-time, cacheable) cost from the per-query execution and reconstruction work, we microbenchmark the planning stages $T_{\mathrm{part}}$ (partitioning) and $T_{\mathrm{gen}}$ (subexperiment-template generation, exact quasiprobability representation $M\!\to\!\infty$) in isolation, with no circuit execution. Table~\ref{tab:planning} reports median times for the ZFeatureMap$+$RealAmplitudes circuit used throughout. Partitioning is cheap and grows only mildly with width ($1.3$\,ms at 4 qubits to $2.4$\,ms at 16 qubits, single cut), so cut \emph{planning} is not the bottleneck and scales well beyond the widths we train on. Template generation tracks the $\mathcal{O}(9^c)$ subexperiment count ($5.5$\,ms, $33$\,ms, $219$\,ms at one, two, three cuts for 12 qubits). Because these artefacts depend only on circuit structure and partition label, they are computed once and cached (Eq.~\eqref{eq:ttotal-amortised}), amortised over the thousands of parameter-binding queries in a run rather than recurring per iteration. The genuine per-query bottleneck is therefore the execution and reconstruction stages, which cannot be amortised because they depend on the live parameters $(\theta,x)$---motivating the restriction to small widths where these stages can be swept systematically.}

\begin{table}[t]
  \centering
  \caption{\revised{Planning-stage microbenchmark (median over repeated trials, no execution). Partitioning $T_{\mathrm{part}}$ scales mildly with circuit width and is independent of cut count; template generation $T_{\mathrm{gen}}$ tracks the $\mathcal{O}(9^c)$ subexperiment count $K$. Both are one-time, cacheable structural costs (Eq.~\eqref{eq:ttotal-amortised}).}}
  \label{tab:planning}
  \small
  \begin{tabular}{cccrrr}
    \toprule
    Qubits & Parts & Cuts & $K$ & $T_{\mathrm{part}}$ (ms) & $T_{\mathrm{gen}}$ (ms) \\
    \midrule
    4  & 2 & 1 & 12  & 1.3 & 2.7 \\
    8  & 2 & 1 & 12  & 1.5 & 4.1 \\
    12 & 2 & 1 & 12  & 2.1 & 5.7 \\
    16 & 2 & 1 & 12  & 2.4 & 7.3 \\
    12 & 3 & 2 & 108 & 2.1 & 33.1 \\
    12 & 4 & 3 & 864 & 2.4 & 218.7 \\
    \bottomrule
  \end{tabular}
\end{table}

This exponential scaling renders systematic \emph{training} experiments beyond 4 qubits computationally prohibitive. We therefore restrict our evaluation to circuits with up to 4 qubits. \revised{This is a scope boundary imposed by the iterative training loop and the requirement for controlled, repeated sweeps---not a claim that circuit cutting itself is limited to this scale.} Our findings about reconstruction dominance and scaling saturation are structural properties that become more pronounced at larger scales.

\textbf{Logging and derived metrics.}
Each run records end-to-end training time, test accuracy, stage-level timings per estimator call, and robustness traces under Gaussian noise and FGSM perturbations. We report reconstruction fractions, scaling ratios, and straggler sensitivity computed only between matched runs (dataset, seed, configuration, and cut label) to ensure observed differences are attributable to the varied parameter.

\subsection{Results and Analysis}

\subsubsection{RQ1: Cutting Overhead}

\begin{figure}[t]
  \centering
  \begin{subfigure}[t]{0.48\columnwidth}
    \centering
    \includegraphics[width=\columnwidth]{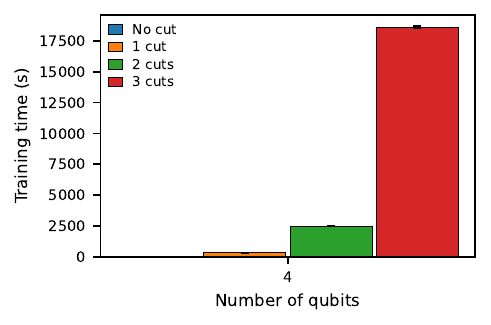}
    \caption{Iris ($\textit{maxiter}=60$)}
    \label{fig:rq1_iris}
  \end{subfigure}\hfill
  \begin{subfigure}[t]{0.48\columnwidth}
    \centering
    \includegraphics[width=\columnwidth]{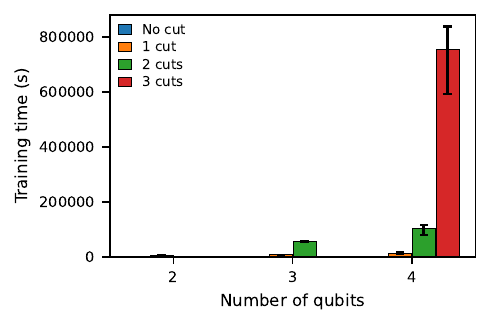}
    \caption{MNIST (10 epochs)}
    \label{fig:rq1_mnist}
  \end{subfigure}
  \caption{\revised{RQ1: End-to-end training time under clean execution.} Training time increases substantially with cut count due to subexperiment expansion and reconstruction overhead.}
  \label{fig:rq1}
\end{figure}

Figure~\ref{fig:rq1} reports end-to-end training time under the clean setting (Iris: $\textit{maxiter}=60$; MNIST: 10 epochs). Training time increases sharply with cut count, consistent with the $\mathcal{O}(9^c)$ subexperiment expansion compounded across all $(q + 2p) \times |\text{batch}|$ estimator queries per iteration. For MNIST, the increase from no cut to three cuts is approximately an order of magnitude ($729\times$ more subexperiments per query plus corresponding reconstruction growth). On Iris, the relative overhead is substantial but lower in absolute terms.

\subsubsection{RQ2: Scalability Bottlenecks}
\label{subsubsec:rq2}

\begin{figure}[t]
  \centering
  \begin{subfigure}[t]{0.48\columnwidth}
    \centering
    \includegraphics[width=\columnwidth]{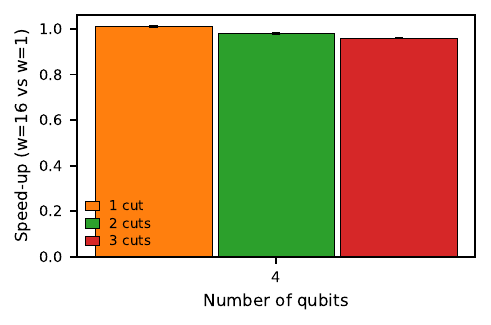}
    \caption{Iris ($\textit{maxiter}=10$)}
    \label{fig:rq2_iris}
  \end{subfigure}\hfill
  \begin{subfigure}[t]{0.48\columnwidth}
    \centering
    \includegraphics[width=\columnwidth]{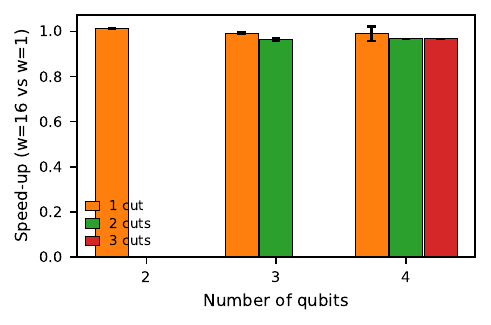}
    \caption{MNIST (5 epochs)}
    \label{fig:rq2_mnist}
  \end{subfigure}
  \caption{\revised{RQ2: Scaling behaviour under clean execution.} Bars report speed-up at 16 workers relative to 1 worker for matched pairs. Speed-up near or below~1 indicates reconstruction overhead dominates.}
  \label{fig:rq2}
\end{figure}

Figure~\ref{fig:rq2} reports speed-up from increasing parallelism (ratio of training time at 1 worker to 16 workers, matched seed--configuration pairs). Speed-ups are close to or below~1 for both datasets, indicating that additional workers yield negligible time reductions.

The worker pool size $w$ controls only concurrent subexperiment execution; it does not affect serial reconstruction. When $T_{\mathrm{rec}}$ dominates, increasing $w$ has diminishing returns---a direct instance of Amdahl's law. Table~\ref{tab:rq2_recfrac} quantifies this: reconstruction accounts for 43\% (1~cut), 49\% (2~cuts), and 53\% (3~cuts, 95th percentile 58\%) of per-query time. Even with perfect execution parallelisation, total query time at three cuts could be reduced by at most 47\%, explaining the scaling saturation in Fig.~\ref{fig:rq2}.

\begin{table}[t]
  \centering
  \caption{RQ2: Reconstruction share ($T_{\mathrm{rec}}/T_{\mathrm{total}}$) from estimator-call summaries under clean execution (\textit{straggler\_delay\_s}$=0.0$). The monotonic increase with cut count confirms that reconstruction progressively dominates per-query time.}
  \label{tab:rq2_recfrac}
  \input{table_rq2_reconstruct_frac.tex}
\end{table}

\subsubsection{RQ3: Straggler Sensitivity}

\begin{figure}[t]
  \centering
  \begin{subfigure}[t]{0.48\columnwidth}
    \centering
    \includegraphics[width=\columnwidth]{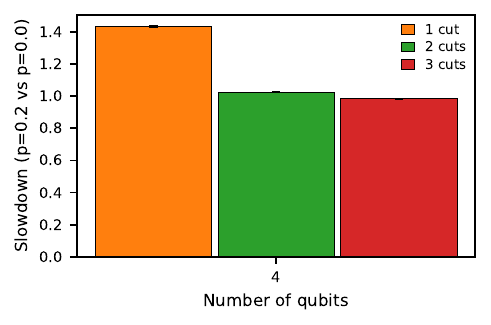}
    \caption{Iris ($\textit{maxiter}=10$, 8 workers)}
    \label{fig:rq3_iris}
  \end{subfigure}\hfill
  \begin{subfigure}[t]{0.48\columnwidth}
    \centering
    \includegraphics[width=\columnwidth]{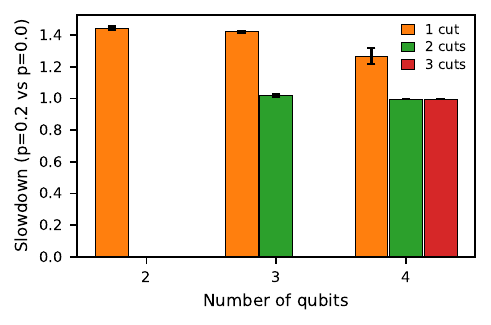}
    \caption{MNIST (3 epochs, 8 workers)}
    \label{fig:rq3_mnist}
  \end{subfigure}
  \caption{\revised{RQ3: Straggler sensitivity at \textit{straggler\_delay\_s}$=0.1$.} Bars report slowdown at $p=0.2$ relative to $p=0.0$. Reconstruction-dominated configurations show lower sensitivity to execution-side stragglers.}
  \label{fig:rq3}
\end{figure}

Figure~\ref{fig:rq3} reports sensitivity to injected stragglers (\textit{straggler\_delay\_s}$=0.1$), showing the slowdown ratio at $p=0.2$ relative to $p=0.0$. Reconstruction-dominated configurations exhibit \emph{lower} relative sensitivity to execution-side stragglers, since $T_{\mathrm{rec}}$ is unaffected by execution delays and the additional straggler time is a smaller fraction of the already-large $T_{\mathrm{total}}$. Conversely, execution-dominated configurations amplify tail latency, because the reconstruction barrier waits for all subexperiments and the slowest straggler determines $T_{\mathrm{exec}}$. Straggler mitigation (speculative execution~\cite{Zaharia2008LATE}, work stealing~\cite{BlumofeLeiserson1999WorkStealing}) is therefore most effective when $T_{\mathrm{exec}}$ dominates; in reconstruction-dominated regimes, effort is better directed at reducing or overlapping reconstruction.

\subsubsection{RQ4: Accuracy}

\begin{figure}[t]
  \centering
  \begin{subfigure}[t]{0.48\columnwidth}
    \centering
    \includegraphics[width=\columnwidth]{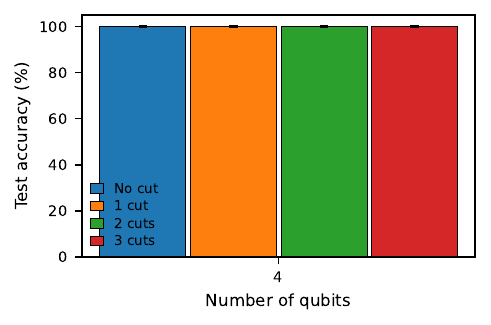}
    \caption{Iris ($\textit{maxiter}=60$)}
    \label{fig:rq4_iris}
  \end{subfigure}\hfill
  \begin{subfigure}[t]{0.48\columnwidth}
    \centering
    \includegraphics[width=\columnwidth]{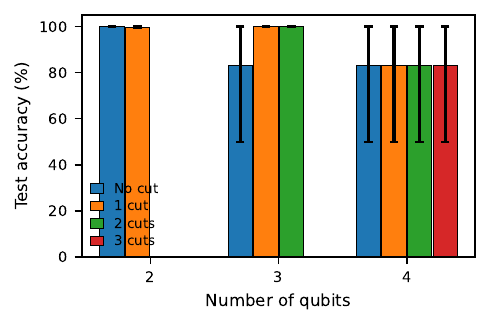}
    \caption{MNIST (10 epochs)}
    \label{fig:rq4_mnist}
  \end{subfigure}
  \caption{\revised{RQ4: Absolute test accuracy under clean execution.} On Iris, accuracy is invariant across cut settings. On MNIST, accuracy is not systematically degraded by cutting.}
  \label{fig:rq4}
\end{figure}

Figure~\ref{fig:rq4} reports test accuracy under the clean setting with the maximum training budgets (Iris: $\textit{maxiter}=60$; MNIST: 10 epochs).

On Iris, all configurations achieve identical test accuracy, confirming that cutting does not degrade predictive performance: reconstruction is mathematically exact up to finite-sample variance, and the low-dimensional landscape is insensitive to the additional estimator noise.

On MNIST, accuracy varies with qubit descriptor and random seed but shows no systematic degradation with cut count: several cut configurations match or exceed the uncut baseline. Cutting changes the effective noise profile of the estimator, which can alter optimisation trajectories and occasionally improve final accuracy through implicit regularisation, even though the expected value of the reconstructed estimate remains unbiased.

\subsubsection{RQ5: Robustness}

\begin{figure}[t]
  \centering
  \begin{subfigure}[t]{0.48\columnwidth}
    \centering
    \includegraphics[width=\columnwidth]{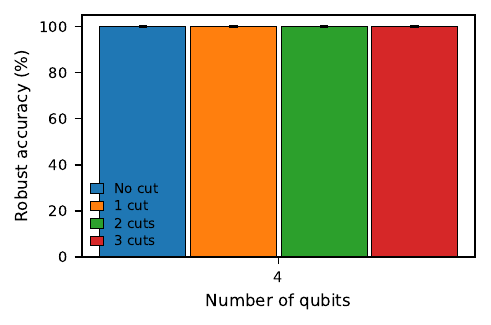}
    \caption{Iris ($\textit{maxiter}=60$)}
    \label{fig:rq5_iris}
  \end{subfigure}\hfill
  \begin{subfigure}[t]{0.48\columnwidth}
    \centering
    \includegraphics[width=\columnwidth]{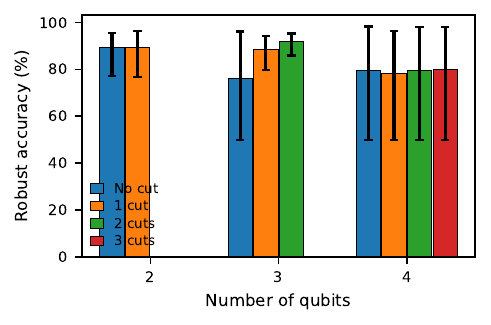}
    \caption{MNIST (10 epochs)}
    \label{fig:rq5_mnist}
  \end{subfigure}
  \caption{\revised{RQ5: Robustness summary under clean execution.} Bars report mean accuracy over non-zero perturbation magnitudes (Gaussian and FGSM). Robustness is not systematically degraded by cutting.}
  \label{fig:rq5}
\end{figure}

We evaluate robustness using perturbation traces under two complementary attack models: additive Gaussian noise (random perturbations to input features) and Fast Gradient Sign Method (FGSM) attacks~\cite{Goodfellow2015Adversarial} (adversarial perturbations along the gradient direction). Figure~\ref{fig:rq5} reports a robustness summary per run, computed as the mean accuracy over non-zero perturbation magnitudes averaged across both attack types.

For Iris, the robustness summary matches clean accuracy across all configurations, consistent with the accuracy invariance in RQ4. For MNIST, robustness varies across configurations, but several cut settings match or exceed the baseline, consistent with the noise-regularisation interpretation: increased estimator variance during training may produce smoother decision boundaries~\cite{Madry2018Robust}. Given the limited seed count, we report these as measured outcomes; a more extensive evaluation would be needed to establish robustness improvement as a reliable effect.

%% file: table_rq2_reconstruct_frac.tex
\begin{tabular}{r r r r}
\toprule
\#cuts & $n$ & median $T_{\mathrm{rec}}/T_{\mathrm{total}}$ & p95 $T_{\mathrm{rec}}/T_{\mathrm{total}}$ \\
\midrule
1 & 41 & 0.430 & 0.474 \\
2 & 28 & 0.489 & 0.533 \\
3 & 5 & 0.530 & 0.580 \\
\bottomrule
\end{tabular}

%% file: section-discussion.tex

%
\section{Validation on Real Quantum Hardware}
\label{sec:validation}

Our main study uses a high-throughput simulator backend so that we can isolate the cut-aware pipeline structure (overheads, scaling, straggler sensitivity) from the confounding latencies of a shared hardware queue, and so that we can afford the large number of repeated estimator queries an iterative training loop demands. This design choice has an important consequence: on a local simulator the per-circuit execution time is very small, so the measured execution stage $T_{\mathrm{exec}}$ is compressed relative to what a real device would incur, and the relative share of reconstruction is correspondingly inflated. To test whether our findings transfer to physical hardware---and to characterise how the stage balance changes in a latency-bound regime---we performed a targeted validation on a real superconducting QPU.

\textbf{Validation protocol.}
Because real-hardware quantum time is a scarce, metered resource, executing the full iterative training loop on a QPU (every forward and parameter-shift query, for every sample and iteration, each expanded by $9^c$) is infeasible within a realistic allocation. We therefore adopt a \emph{train-on-simulator, evaluate-on-hardware} protocol: model weights are trained to convergence on the statevector simulator (which, as established in Sections~\ref{sec:system-model}--\ref{sec:evaluation}, produces gradients mathematically equivalent to the uncut model up to finite-sample variance), and the trained model is then evaluated on the QPU under each cut configuration. This isolates the quantity of interest---the \emph{execution} of cut-induced subexperiments and their classical reconstruction on real hardware---while keeping the hardware budget tractable. We evaluate the Iris binary task with a 4-qubit ZFeatureMap$+$RealAmplitudes circuit on \texttt{ibm\_kingston} (IBM~Heron~r2, 156 qubits), with 1024 shots per subexperiment, circuits transpiled at optimisation level~1, and reconstruction performed on the classical host of Section~\ref{subsec:setup}. We report clean test accuracy and robustness (Gaussian and FGSM at magnitude $0.1$) on both the simulator and the QPU, alongside the hardware stage timings.

\textbf{Findings.}
Table~\ref{tab:qpu} summarises the results. Three observations stand out.
\emph{(i)~Functional correctness on hardware.} Reconstruction from physically measured subexperiment outcomes reproduces the correct predictions: QPU clean accuracy matches the simulator (100\% for every configuration on this task), and robustness under both Gaussian and FGSM perturbations is likewise preserved (100\%). Circuit cutting therefore composes correctly with real-device noise at this scale---the decomposition and quasiprobability reconstruction are not silently corrupted by hardware error.
\emph{(ii)~The stage balance inverts on hardware.} Whereas on the simulator execution is near-instantaneous, on the QPU the execution stage becomes the heavy stage---dominated by per-circuit device time, transpilation, and queueing latency---and grows sharply with the subexperiment count: the number of executed circuits scales with the $\mathcal{O}(9^c)$ overhead (from the uncut baseline to one cut and beyond), and the measured execution wall-time and on-device run-time rise accordingly, while the classical reconstruction time remains small (tens to a few hundred milliseconds). Subexperiments here were submitted sequentially in chunks to a single backend, so the reported $T_{\mathrm{exec}}$ reflects single-device execution plus queueing; dispatching the independent subexperiments across parallel QPUs would reduce this term---a direct hardware realisation of the worker-pool model that we leave to future work.
\emph{(iii)~The two regimes are complementary, and our architectural conclusion is unchanged.} The $9^c$ growth inflates \emph{both} stages---more subexperiments to execute and more coefficient-weighted product terms to reconstruct. The simulator isolates the reconstruction growth (a purely classical cost that better quantum hardware cannot reduce); the hardware exposes the execution-latency growth (which better hardware and queueing can reduce). The central message therefore holds under either backend: cutting transforms training into a staged distributed workload whose critical path is shaped by an exponentially growing subexperiment set, and whichever stage dominates, the bottleneck is a systems-level property of the expansion rather than an artefact of the backend.

\textbf{Incorporating realistic execution delays.}
The methodology accommodates real-world execution cost in two ways without requiring full-loop hardware runs. First, the synthetic straggler-injection mechanism of Section~\ref{sec:methodology} already perturbs $T_{\mathrm{exec}}$ at the task level; the per-circuit run-times measured here (Table~\ref{tab:qpu}) and historical device-execution statistics published by QPU providers can parameterise the injected-delay distribution, yielding a hardware-calibrated execution model on top of the simulator pipeline. Second, the amortisation analysis of Eq.~\eqref{eq:ttotal-amortised} continues to hold on hardware: the structural stages are computed once and cached, so the recurring per-iteration cost is the parameter-binding, execution, and reconstruction terms, exactly the quantities the hardware validation measures.

\textbf{Scope.}
This validation is deliberately small: it is an inference-time evaluation under fixed weights on a limited test set and a single backend, and it does not perform hardware-based gradient training. Accordingly, the accuracy- and robustness-preservation conclusions of RQ4--RQ5 remain established on the simulator; the hardware experiment confirms functional correctness under real device noise and characterises the execution/reconstruction balance in a latency-bound regime, rather than providing a full hardware scaling study, which we identify as future work (Section~\ref{sec:discussion}).

\begin{table}[t]
  \centering
  \caption{Real-hardware validation on \texttt{ibm\_kingston} (IBM~Heron~r2, 156 qubits), Iris binary task, 4 qubits, 1024 shots, train-on-simulator/evaluate-on-hardware. $K$ is the number of executed subexperiment circuits per evaluation round; $T_{\mathrm{exec}}^{\mathrm{wall}}$ includes queueing and chunked submission, $t_{\mathrm{run}}$ is on-device execution time, $T_{\mathrm{rec}}$ is classical reconstruction. Accuracy and robustness are identical on simulator and QPU.}
  \label{tab:qpu}
  \setlength{\tabcolsep}{4pt}
  \resizebox{\columnwidth}{!}{%
  \begin{tabular}{lrrrrr}
    \toprule
    Config & $K$ & $T_{\mathrm{exec}}^{\mathrm{wall}}$ (s) & $t_{\mathrm{run}}$ (s) & $T_{\mathrm{rec}}$ (s) & QPU Acc. \\
    \midrule
    No cut (AAAA) & 10  & 10.73 & 0.00 & 0.028 & 100\% \\
    1 cut (AABB)  & 120 & 52.52 & 9.27 & 0.195 & 100\% \\
    2 cuts (AABC) & 1080 & 458.84 & 367.45 & 1.676 & 100\% \\
    \bottomrule
  \end{tabular}}
\end{table}

\section{Discussion and Future Work}
\label{sec:discussion}

Our evaluation demonstrates that circuit cutting should be treated as a staged distributed workload rather than a purely algorithmic transformation. The central observation across RQ1--RQ3 is that, while cutting increases parallelisable work, it simultaneously introduces substantial non-parallel components---particularly reconstruction---that dominate end-to-end performance. This section consolidates these findings, discusses the practical implications, and outlines future directions.

\subsection{Discussion}

\textbf{Reconstruction is the primary scaling limiter.}
RQ2 quantifies reconstruction as a substantial and growing share of per-query time (53\% median at three cuts), inducing two scaling constraints. First, the serial reconstruction stage bounds achievable speed-up even under ideal execution parallelism---a direct instance of Amdahl's law applied to the cut-aware pipeline. \revised{Amdahl's law~\cite{Amdahl1967} states that if a fraction $f$ of a workload is inherently serial and the remaining $(1-f)$ is perfectly parallelisable across $w$ workers, the overall speed-up is bounded by $1/(f + (1-f)/w)$, which converges to $1/f$ as $w \to \infty$; here the serial fraction $f$ is the reconstruction barrier, which cannot be overlapped with execution in the unoptimised pipeline.} At $f = 0.53$ (three cuts, median), this yields a maximum theoretical speed-up of approximately $1.89\times$, regardless of how many workers are available. Second, reconstruction acts as a global barrier: the training loop does not progress until all subexperiment results are available, making the pipeline sensitive to upstream tail latency. These effects explain the speed-up saturation observed in Fig.~\ref{fig:rq2} and indicate that reducing the reconstruction fraction is a prerequisite for meaningful parallel scaling.

\revised{%
\textbf{Reconstruction can be partially overlapped and parallelised.}
Our current implementation treats reconstruction as a single serial reduction that begins only after a global barrier on all subexperiment results, which is the behaviour measured throughout RQ2. However, the structure of Eq.~\eqref{eq:reconstruction} shows that this barrier is not fundamental. Because the target estimate is a sum over coefficient-weighted tensor products of \emph{per-fragment} outcomes, two forms of relaxation are feasible. (i)~\emph{Overlap with execution:} partial products can be accumulated incrementally as fragment results arrive, so reconstruction of the terms whose fragments have already completed can proceed while other subexperiments are still executing, hiding part of $T_{\mathrm{rec}}$ behind $T_{\mathrm{exec}}$ rather than serialising after it. (ii)~\emph{Intra-reconstruction parallelism:} the $\mathcal{O}(9^c)$ product terms in Eq.~\eqref{eq:reconstruction} are mutually independent and can be evaluated concurrently and combined by a parallel (tree) reduction, lowering the span of the summation from linear to logarithmic in the number of terms. Neither relaxation reduces the exponential \emph{work} of reconstruction---that is intrinsic to the decomposition---but both reduce its contribution to the critical path and therefore relax the Amdahl bound above by shrinking the effective serial fraction $f$. We did not implement these optimisations in the present study because our goal was to \emph{measure} the unoptimised pipeline and establish reconstruction as the bottleneck; realising overlapped and tree-reduced reconstruction is a concrete and, we expect, high-value engineering direction, which we identify as future work below.}

\textbf{The $\mathcal{O}(9^c)$ overhead compounds across training.}
The exponential overhead is incurred once per estimator query, not once per run. Including the $2p$ parameter-shift gradient queries, the total subexperiments per epoch scale as $\mathcal{O}(9^c \cdot p \cdot |\text{dataset}|)$---the primary reason training time increases by an order of magnitude at three cuts.

\textbf{Straggler sensitivity is modulated by critical-path structure.}
RQ3 shows that reconstruction-dominated configurations exhibit lower sensitivity to execution-side delays, while execution-dominated configurations amplify tail latency. The optimal mitigation strategy therefore depends on where the bottleneck lies: classical straggler mitigation in execution-dominated regimes, reconstruction optimisation in reconstruction-dominated regimes.

\textbf{Learning outcomes are preserved under cutting.}
RQ4 confirms that accuracy is fully preserved on Iris and not systematically degraded on MNIST. RQ5 demonstrates that robustness under Gaussian noise and FGSM perturbations is maintained, with some cut settings exhibiting comparable or improved robustness. These findings validate that circuit cutting can serve as a systems-level execution strategy without silently degrading model quality or creating exploitable vulnerabilities in decision boundaries.

\textbf{Implications for system design.}
Our findings suggest that: (i)~reconstruction should be treated as a first-class optimisation target; (ii)~scheduling policies should adapt to the reconstruction fraction; (iii)~resource allocation should balance execution workers and reconstruction compute; and (iv)~the exponential overhead profile should inform decisions about when cutting is preferable to alternatives (model compression, shallower circuits, or waiting for larger devices).

\subsection{Scope and Generalisability}

We identify the following scope boundaries that contextualise our results.

\revised{\textbf{Experimental scale.} The $\mathcal{O}(9^c)$ growth of subexperiment counts restricts practical experimentation to small qubit counts when cutting is embedded in an iterative training loop (Section~\ref{subsec:setup}). Single-shot cutting toolflows such as CutQC target much larger circuits~\cite{Tang2021CutQC}; our scope limit stems from the multiplicative interaction of cutting overhead with the training-query volume, not from cutting per se. The structural findings---reconstruction dominance, scaling saturation, and straggler sensitivity patterns---are architectural properties of the pipeline that become more pronounced at larger scales, since the reconstruction share increases monotonically with cut count (Table~\ref{tab:rq2_recfrac}).}

\textbf{Runtime and evaluation scope.} Our thread-based worker pool isolates pipeline effects from network confounders; the qualitative findings are structural properties rather than threading artefacts. The number of independent runs per configuration (notably $n=5$ at 3~cuts) limits statistical power; we report medians and 95th percentiles accordingly. The observed trends are directionally consistent across all configurations and both datasets.

\subsection{Future Work}

\textbf{Minimising reconstruction overhead.}
The most pressing direction is reducing the reconstruction bottleneck. Promising approaches include approximate reconstruction via tensor-network contractions or low-rank methods, distributed tree-reduction to parallelise the reconstruction phase, incremental reconstruction overlapping with late execution, and integration of reduced-overhead cutting schemes~\cite{Lowe2023FastCutting,Piveteau2022Knitting} that reduce the base of the exponential from 9 to smaller constants.

\textbf{Scheduling and extended evaluation.}
A systematic comparison of batched and staggered dispatch strategies, variance-aware schedulers, and adaptive shot allocation would clarify how scheduling policies interact with the reconstruction barrier. Validating the key findings on a distributed runtime (e.g., Ray~\cite{Moritz2018Ray}), with additional datasets, ansatz depths, multi-class problems, and detailed convergence analysis would strengthen generalisability.

%% file: section-conclusion.tex

\section{Conclusions}
\label{sec:conclusion}

We proposed \distest{}, a cut-aware estimator execution pipeline for training quantum neural networks via circuit cutting, and evaluated it as a staged distributed workload with explicit measurement of overheads, bottlenecks, and learning outcomes.

Our results demonstrate that circuit cutting introduces substantial end-to-end overheads that grow with cut count (RQ1). Classical reconstruction is a primary contributor to wall-clock time, reaching a median of 53\% and a 95th percentile of 58\% of per-query estimator time at three cuts, thereby limiting achievable speed-up under increased parallelism (RQ2). Straggler-induced tail effects measurably impact time-to-solution, with sensitivity governed by the balance between execution and reconstruction on the critical path (RQ3). Despite these systems costs, test accuracy is fully preserved on Iris and maintained without systematic degradation on MNIST across all evaluated cut configurations (RQ4). Robustness under Gaussian noise and FGSM perturbations is similarly preserved, with several cut configurations exhibiting comparable or improved robustness relative to the uncut baseline (RQ5).

The exponential growth of subexperiment counts with each additional cut ($\mathcal{O}(9^c)$ for CNOT-based wire cutting) represents a fundamental computational barrier that constrains practical experimentation to small qubit counts with current methods. \revised{A targeted validation on the \texttt{ibm\_kingston} superconducting QPU confirms that reconstruction from physically measured subexperiment outcomes preserves both accuracy and robustness, while revealing that the stage balance inverts on real hardware: execution becomes the dominant cost due to per-circuit device time and queueing, whereas reconstruction stays small in absolute terms. The central architectural conclusion is unchanged under either regime, because the $\mathcal{O}(9^c)$ growth inflates both stages---reconstruction is a purely classical cost that better hardware cannot reduce, and execution latency is a hardware cost that reconstruction optimisation cannot reduce.}

This motivates a focused effort on overhead reduction---through approximate reconstruction, distributed reduction, and reduced-overhead cutting schemes---as a prerequisite for scaling circuit cutting to practically relevant QML workloads. Overall, our study establishes that scalable circuit cutting for learning workloads requires treating reconstruction and aggregation as first-class systems problems, and provides a reproducible foundation for designing the scheduling, reconstruction, and variance-aware execution strategies needed to make cutting practical at scale. In addition to the future directions discussed previously, we will explore the integration of our \distest{} in quantum application deployment and management frameworks such as QFaaS~\cite{qfaas}.

%% file: section-credit.tex
\section*{CRediT Authorship Contribution Statement}
\label{sec:credit}
\textbf{Prabhjot Singh:} Conceptualization, Methodology, Software, Investigation, Data curation, Formal analysis, Visualization, Writing -- original draft.
\textbf{Adel N. Toosi:} Conceptualization, Supervision, Writing -- review \& editing.
\textbf{Rajkumar Buyya:} Conceptualization, Supervision, Methodology, Project administration, Writing -- review \& editing.